\documentclass[twocolumn]{aastex631}

\shorttitle{Direct Optimal Mapping Image Power Spectrum}
\shortauthors{Xu et al.}

\usepackage{CJK}

\graphicspath{{./}{figures/}}

\newcommand{\dg}{$^\circ$}

\defcitealias{xu/etal:2022}{X22}

\begin{document}
\begin{CJK*}{UTF8}{gbsn}

\title{Direct Optimal Mapping Image Power Spectrum and its Window Functions}

\correspondingauthor{Zhilei Xu (徐智磊)}
\email{zhileixu@mit.edu\\zhileixu@space.mit.edu}

\author[0000-0001-5112-2567]{Zhilei  Xu}
\affiliation{MIT Kavli Institute, Massachusetts Institute of Technology, Cambridge, MA}

\author[0000-0001-5421-8927]{Honggeun  Kim}
\affiliation{MIT Kavli Institute, Massachusetts Institute of Technology, Cambridge, MA}
\affiliation{Department of Physics, Massachusetts Institute of Technology, Cambridge, MA}

\author[0000-0002-4117-570X]{Jacqueline N. Hewitt}
\affiliation{MIT Kavli Institute, Massachusetts Institute of Technology, Cambridge, MA}
\affiliation{Department of Physics, Massachusetts Institute of Technology, Cambridge, MA}

\author[0000-0002-3839-0230]{Kai-Feng  Chen}
\affiliation{MIT Kavli Institute, Massachusetts Institute of Technology, Cambridge, MA}
\affiliation{Department of Physics, Massachusetts Institute of Technology, Cambridge, MA}

\author[0000-0002-8211-1892]{Nicholas S. Kern}
\affiliation{MIT Kavli Institute, Massachusetts Institute of Technology, Cambridge, MA}
\affiliation{Department of Physics, Massachusetts Institute of Technology, Cambridge, MA}
\affiliation{NHFP Hubble Fellow}

\author{Eleanor  Rath}
\affiliation{MIT Kavli Institute, Massachusetts Institute of Technology, Cambridge, MA}
\affiliation{Department of Physics, Massachusetts Institute of Technology, Cambridge, MA}

\author{Ruby  Byrne}
\affiliation{Cahill Center for Astronomy and Astrophysics, California Institute of Technology, Pasadena, CA}

\author{Ad\'elie  Gorce}
\affiliation{Department of Physics and McGill Space Institute, McGill University, 3600 University Street, Montreal, QC H3A 2T8, Canada}

\author[0000-0003-0073-5528]{Robert Pascua}
\affiliation{Department of Physics and McGill Space Institute, McGill University, 3600 University Street, Montreal, QC H3A 2T8, Canada}

\author{Zachary E. Martinot}
\affiliation{Department of Physics and Astronomy, University of Pennsylvania, Philadelphia, PA}

\author[0000-0003-3336-9958]{Joshua S. Dillon}
\affiliation{Department of Astronomy, University of California, Berkeley, CA}

\author[0000-0001-7532-645X]{Bryna J. Hazelton}
\affiliation{Department of Physics, University of Washington, Seattle, WA}
\affiliation{eScience Institute, University of Washington, Seattle, WA}

\author[0000-0001-6876-0928]{Adrian  Liu}
\affiliation{Department of Astronomy, University of California, Berkeley, CA}
\affiliation{Department of Physics and McGill Space Institute, McGill University, 3600 University Street, Montreal, QC H3A 2T8, Canada}

\author[0000-0001-7694-4030]{Miguel F. Morales}
\affiliation{Department of Physics, University of Washington, Seattle, WA}

\author{Zara  Abdurashidova}
\affiliation{Department of Astronomy, University of California, Berkeley, CA}

\author{Tyrone  Adams}
\affiliation{South African Radio Astronomy Observatory, Black River Park, 2 Fir Street, Observatory, Cape Town, 7925, South Africa}

\author[0000-0002-4810-666X]{James E. Aguirre}
\affiliation{Department of Physics and Astronomy, University of Pennsylvania, Philadelphia, PA}

\author{Paul  Alexander}
\affiliation{Cavendish Astrophysics, University of Cambridge, Cambridge, UK}

\author{Zaki S. Ali}
\affiliation{Department of Astronomy, University of California, Berkeley, CA}

\author{Rushelle  Baartman}
\affiliation{South African Radio Astronomy Observatory, Black River Park, 2 Fir Street, Observatory, Cape Town, 7925, South Africa}

\author{Yanga  Balfour}
\affiliation{South African Radio Astronomy Observatory, Black River Park, 2 Fir Street, Observatory, Cape Town, 7925, South Africa}

\author[0000-0001-9428-8233]{Adam P. Beardsley}
\affiliation{School of Earth and Space Exploration, Arizona State University, Tempe, AZ}
\affiliation{Department of Physics, Winona State University, Winona, MN}

\author[0000-0002-0916-7443]{Gianni  Bernardi}
\affiliation{INAF-Istituto di Radioastronomia, via Gobetti 101, 40129 Bologna, Italy}
\affiliation{Department of Physics and Electronics, Rhodes University, PO Box 94, Grahamstown, 6140, South Africa}
\affiliation{South African Radio Astronomy Observatory, Black River Park, 2 Fir Street, Observatory, Cape Town, 7925, South Africa}

\author{Tashalee S. Billings}
\affiliation{Department of Physics and Astronomy, University of Pennsylvania, Philadelphia, PA}

\author[0000-0002-8475-2036]{Judd D. Bowman}
\affiliation{School of Earth and Space Exploration, Arizona State University, Tempe, AZ}

\author{Richard F. Bradley}
\affiliation{National Radio Astronomy Observatory, Charlottesville, VA}

\author[0000-0001-5668-3101]{Philip  Bull}
\affiliation{Jodrell Bank Centre for Astrophysics, University of Manchester, Manchester, M13 9PL, United Kingdom}
\affiliation{Department of Physics and Astronomy,  University of Western Cape, Cape Town, 7535, South Africa}

\author[0000-0002-8465-9341]{Jacob  Burba}
\affiliation{Jodrell Bank Centre for Astrophysics, University of Manchester, Manchester, M13 9PL, United Kingdom}

\author{Steven  Carey}
\affiliation{Cavendish Astrophysics, University of Cambridge, Cambridge, UK}

\author[0000-0001-6647-3861]{Chris L. Carilli}
\affiliation{National Radio Astronomy Observatory, Socorro, NM 87801, USA}

\author{Carina  Cheng}
\affiliation{Department of Astronomy, University of California, Berkeley, CA}

\author[0000-0003-3197-2294]{David R. DeBoer}
\affiliation{Radio Astronomy Lab, University of California, Berkeley, CA}

\author{Eloy  de~Lera~Acedo}
\affiliation{Cavendish Astrophysics, University of Cambridge, Cambridge, UK}

\author{Matt  Dexter}
\affiliation{Radio Astronomy Lab, University of California, Berkeley, CA}

\author{Nico  Eksteen}
\affiliation{South African Radio Astronomy Observatory, Black River Park, 2 Fir Street, Observatory, Cape Town, 7925, South Africa}

\author{John  Ely}
\affiliation{Cavendish Astrophysics, University of Cambridge, Cambridge, UK}

\author[0000-0002-0086-7363]{Aaron  Ewall-Wice}
\affiliation{Department of Astronomy, University of California, Berkeley, CA}
\affiliation{Department of Physics, University of California, Berkeley, CA}

\author{Nicolas  Fagnoni}
\affiliation{Cavendish Astrophysics, University of Cambridge, Cambridge, UK}

\author{Randall  Fritz}
\affiliation{South African Radio Astronomy Observatory, Black River Park, 2 Fir Street, Observatory, Cape Town, 7925, South Africa}

\author[0000-0002-0658-1243]{Steven R. Furlanetto}
\affiliation{Department of Physics and Astronomy, University of California, Los Angeles, CA}

\author{Kingsley  Gale-Sides}
\affiliation{Cavendish Astrophysics, University of Cambridge, Cambridge, UK}

\author{Brian  Glendenning}
\affiliation{National Radio Astronomy Observatory, Socorro, NM}

\author[0000-0002-0829-167X]{Deepthi  Gorthi}
\affiliation{Department of Astronomy, University of California, Berkeley, CA}

\author[0000-0002-4085-2094]{Bradley  Greig}
\affiliation{School of Physics, University of Melbourne, Parkville, VIC 3010, Australia}

\author{Jasper  Grobbelaar}
\affiliation{South African Radio Astronomy Observatory, Black River Park, 2 Fir Street, Observatory, Cape Town, 7925, South Africa}

\author{Ziyaad  Halday}
\affiliation{South African Radio Astronomy Observatory, Black River Park, 2 Fir Street, Observatory, Cape Town, 7925, South Africa}

\author{Jack  Hickish}
\affiliation{Radio Astronomy Lab, University of California, Berkeley, CA}

\author[0000-0002-0917-2269]{Daniel C. Jacobs}
\affiliation{School of Earth and Space Exploration, Arizona State University, Tempe, AZ}

\author{Austin  Julius}
\affiliation{South African Radio Astronomy Observatory, Black River Park, 2 Fir Street, Observatory, Cape Town, 7925, South Africa}

\author{MacCalvin  Kariseb}
\affiliation{South African Radio Astronomy Observatory, Black River Park, 2 Fir Street, Observatory, Cape Town, 7925, South Africa}

\author[0000-0002-1876-272X]{Joshua  Kerrigan}
\affiliation{Department of Physics, Brown University, Providence, RI}

\author[0000-0003-0953-313X]{Piyanat  Kittiwisit}
\affiliation{Department of Physics and Astronomy,  University of Western Cape, Cape Town, 7535, South Africa}

\author[0000-0001-6744-5328]{Saul A. Kohn}
\affiliation{Department of Physics and Astronomy, University of Pennsylvania, Philadelphia, PA}

\author[0000-0002-2950-2974]{Matthew  Kolopanis}
\affiliation{School of Earth and Space Exploration, Arizona State University, Tempe, AZ}

\author{Adam  Lanman}
\affiliation{Department of Physics, Brown University, Providence, RI}

\author[0000-0002-4693-0102]{Paul  La~Plante}
\affiliation{Department of Astronomy, University of California, Berkeley, CA}
\affiliation{Department of Physics and Astronomy, University of Pennsylvania, Philadelphia, PA}

\author{Anita  Loots}
\affiliation{South African Radio Astronomy Observatory, Black River Park, 2 Fir Street, Observatory, Cape Town, 7925, South Africa}

\author{David Harold~Edward MacMahon}
\affiliation{Radio Astronomy Lab, University of California, Berkeley, CA}

\author{Lourence  Malan}
\affiliation{South African Radio Astronomy Observatory, Black River Park, 2 Fir Street, Observatory, Cape Town, 7925, South Africa}

\author{Cresshim  Malgas}
\affiliation{South African Radio Astronomy Observatory, Black River Park, 2 Fir Street, Observatory, Cape Town, 7925, South Africa}

\author{Keith  Malgas}
\affiliation{South African Radio Astronomy Observatory, Black River Park, 2 Fir Street, Observatory, Cape Town, 7925, South Africa}

\author{Bradley  Marero}
\affiliation{South African Radio Astronomy Observatory, Black River Park, 2 Fir Street, Observatory, Cape Town, 7925, South Africa}

\author[0000-0003-3374-1772]{Andrei  Mesinger}
\affiliation{Scuola Normale Superiore, 56126 Pisa, PI, Italy}

\author{Mathakane  Molewa}
\affiliation{South African Radio Astronomy Observatory, Black River Park, 2 Fir Street, Observatory, Cape Town, 7925, South Africa}

\author{Tshegofalang  Mosiane}
\affiliation{South African Radio Astronomy Observatory, Black River Park, 2 Fir Street, Observatory, Cape Town, 7925, South Africa}

\author[0000-0003-3059-3823]{Steven G. Murray}
\affiliation{School of Earth and Space Exploration, Arizona State University, Tempe, AZ}

\author[0000-0001-7776-7240]{Abraham R. Neben}
\affiliation{MIT Kavli Institute, Massachusetts Institute of Technology, Cambridge, MA}
\affiliation{Department of Physics, Massachusetts Institute of Technology, Cambridge, MA}

\author{Bojan  Nikolic}
\affiliation{Cavendish Astrophysics, University of Cambridge, Cambridge, UK}

\author{Hans  Nuwegeld}
\affiliation{South African Radio Astronomy Observatory, Black River Park, 2 Fir Street, Observatory, Cape Town, 7925, South Africa}

\author[0000-0002-5400-8097]{Aaron R. Parsons}
\affiliation{Department of Astronomy, University of California, Berkeley, CA}

\author[0000-0002-9457-1941]{Nipanjana  Patra}
\affiliation{Department of Astronomy, University of California, Berkeley, CA}

\author{Samantha  Pieterse}
\affiliation{South African Radio Astronomy Observatory, Black River Park, 2 Fir Street, Observatory, Cape Town, 7925, South Africa}

\author{Nima  Razavi-Ghods}
\affiliation{Cavendish Astrophysics, University of Cambridge, Cambridge, UK}

\author{James  Robnett}
\affiliation{National Radio Astronomy Observatory, Socorro, NM 87801, USA}

\author{Kathryn  Rosie}
\affiliation{South African Radio Astronomy Observatory, Black River Park, 2 Fir Street, Observatory, Cape Town, 7925, South Africa}

\author[0000-0002-2871-0413]{Peter  Sims}
\affiliation{Department of Physics and McGill Space Institute, McGill University, 3600 University Street, Montreal, QC H3A 2T8, Canada}

\author{Craig  Smith}
\affiliation{South African Radio Astronomy Observatory, Black River Park, 2 Fir Street, Observatory, Cape Town, 7925, South Africa}

\author{Hilton  Swarts}
\affiliation{South African Radio Astronomy Observatory, Black River Park, 2 Fir Street, Observatory, Cape Town, 7925, South Africa}

\author{Jianrong Tan}
\affiliation{Department of Physics and Astronomy, University of Pennsylvania, Philadelphia, PA}

\author[0000-0003-1602-7868]{Nithyanandan  Thyagarajan}
\affiliation{Commonwealth Scientific and Industrial Research Organisation (CSIRO), Space \& Astronomy, P. O. Box 1130, Bentley, WA 6102, Australia}

\author{Pieter  van~Wyngaarden}
\affiliation{South African Radio Astronomy Observatory, Black River Park, 2 Fir Street, Observatory, Cape Town, 7925, South Africa}

\author[0000-0003-3734-3587]{Peter K.~G. Williams}
\affiliation{Center for Astrophysics, Harvard \& Smithsonian, Cambridge, MA}
\affiliation{American Astronomical Society, Washington, DC}

\author{Haoxuan  Zheng}
\affiliation{Department of Physics, Massachusetts Institute of Technology, Cambridge, MA}

\collaboration{83}{(HERA Collaboration)}

\begin{abstract}
The key to detecting neutral hydrogen during the epoch of reionization (EoR) is to separate the cosmological signal from the dominating foreground radiation.
We developed direct optimal mapping (DOM) to map interferometric visibilities; it contains only linear operations, with full knowledge of point spread functions from visibilities to images.
Here, we demonstrate a fast Fourier transform-based image power spectrum and its window functions computed from the direct optimal mapping images.
We use noiseless simulation, based on the Hydrogen Epoch of Reionization Array Phase I configuration, to study the image power spectrum properties.
The window functions show $<10^{-11}$ of the integrated power leaks from the foreground-dominated region into the EoR window; the 2D and 1D power spectra also verify the separation between the foregrounds and the EoR.

\end{abstract}



\section{Introduction} \label{sec:intro}
Recent cosmological observations have established the standard cosmological model --- $\Lambda$CDM cosmology~\citep{bennett/etal:1996, riess/etal:1998, bennett/etal:2013, hinshaw/etal:2013, planck/etal:2016, planck/etal:2020}.
The cosmic microwave background (CMB) measures the early universe at redshift $\sim$1100~\citep{fixsen/etal:1996, hu/dodelson:2002, staggs/dunkley/page:2018}; late universe measurement gives statistical properties of the universe below redshift 10~\citep{anderson/etal:2014, alam/etal:2017, des:2022, surhud/etal:2023}.
However, we have hardly observed the universe from redshift 1100 to 10, including dark ages, cosmic dawn, and the epoch of reionization (EoR)~\citep{furlanetto/oh/briggs:2006, pritchard/loeb:2012}.

With measurements from the initial ($z\sim1100$) and final ($z<10$) conditions, the $\Lambda$CDM cosmology describes the initial perturbations' growth from linear to non-linear until the ignition of first stars and galaxies~\citep{hogan/rees:1979, madau/meiksin/rees:1997}.
Our goal is to observe this process by measuring radiation from the dominant baryonic content in the universe --- neutral hydrogen.
The atomic hyperfine structure of neutral hydrogen leads to emission or absorption of 21\,cm radiation in the restframe.
The 21\,cm radiation, redshifted, is observed at different frequency channels, tracing the distribution of the baryonic matter during EoR, cosmic dawn, and dark ages~\citep{pritchard/loeb:2012}.
Direct observations covering the above epochs are the ultimate goal of 21\,cm cosmology.
The 21\,cm observations have the potential to measure a wide range of cosmological history, testing the current cosmological model~\citep{mao/etal:2008}.
The result will reveal more details of the cosmological evolution and the universe's contents, including dark matter and dark energy.
The frequency coverage provides a tomographic measurement of the Epoch of Reionization (EoR), constraining the evolution of neutral hydrogen spin temperature and ionization fraction during EoR~\citep{furlanetto/oh/briggs:2006, morales/wyithe:2010, pritchard/loeb:2012, liu/shaw:2020}.
These measurements also constrain sources for the reionization process, including UV and X-ray properties of high-redshift galaxies~\citep{ewall-wice/etal:2016, greig/mesinger/pober:2016, kern/etal:2017}.

In practice, wavelengths of high-redshift 21\,cm radiation are redshifted to $>$\,2\,m beyond $z \sim 8.5$; correspondingly,
$>$\,230\,m baselines are required to achieve $\sim0.5$\dg{} angular resolutions.
Therefore, high-redshift 21\,cm measurements are conducted with interferometers for angular resolutions~\citep{parsons/etal:2010, tingay/etal:2013, vanhaarlem/etal:2013, deboer/etal:2017}.
Interferometers are also cost-effective in achieving large collecting areas, which eventually determines the measurement sensitivity.
We aim to measure the distribution of neutral hydrogen radiation with power spectrum analysis.
Power spectrum is a powerful tool for containing cosmological models as has been demonstrated with CMB~\citep{bennett/etal:2013, planck/etal:2020}, galaxy surveys~\citep{chan/etal:2022, desi:2024}, and other 21\,cm experiments~\citep{dillon/etal:2014, dillon/etal:2015b, beardsley/etal:2016, trott/etal:2016, patil/etal:2017, Li/etal:2019, mertens/etal:2020, rahimi/etal:2021}.

In previous work, we have developed a new mapping algorithm called direct optimal mapping (DOM)~\citep[][hereafter \citetalias{xu/etal:2022}]{xu/etal:2022}.
DOM provides an optimal algorithm to map visibilities to images with only linear operations; the linear operation provides full knowledge of the point spread function and full knowledge of the covariance matrix among pixels.
Since DOM is not based on the Fourier transform, the emission of pixels is estimated independently; analysts are free to choose only the pixels of interest, even from disjoint sky patches.
DOM treats pixels with point sources and extended emission equally, which is critical for precision cosmology that focuses on diffuse emission. 
DOM does not grid visibilities~\citep{barry/chokshi:2022}; it naturally includes the $w$-term in the calculation, so the configuration of the antennas need not be coplanar.
Next-generation radio interferometers are proposed to contain thousands of antennas~\citep{hallinan/etal:2019, slosar/etal:2019}, storing raw visibilities may not be feasible given the $N^2$-scaling; instead storing the image product is a feasible solution.
Future instruments can integrate the imaging calculation into hardware so that the output of the instruments is directly in images, and DOM provides a solution for that with only linear operations.
Designed as an optimal mapping algorithm, DOM converts the visibilities to sky images without losing cosmological information~\citep{tegmark:1997, dillon/etal:2015}.
We refer interested readers to \citetalias{xu/etal:2022} for more details about DOM.

In this paper, we present an image power spectrum based on DOM.
The image power spectrum serves as a consistency check of the DOM algorithm and also provides
a new power spectrum estimator that may be used to verify power spectra estimated with other techniques.
In addition, the image power spectrum technique, if it enables more coherent averaging of visibilities, is potentially more sensitive than delay spectrum techniques that usually incorporate some incoherent averaging~\citep{morales/etal:2019}.
Image power spectra with dense $uv$ sampling also have the potential to remove the foreground `wedge' that limits the useable region of Fourier space when estimating the power spectrum.
Here, we apply the traceability of the DOM's linear operations and we calculate the
exact power spectrum window functions for HERA data.
The window function includes both the mapping step and the power spectrum estimation step.
Section~\ref{sec:image_ps} reviews the available power spectrum estimators, including the delay power spectrum and three image power spectrum estimators.
Section~\ref{sec:dom_ps} introduces the DOM image power spectrum and its window functions.
In Section~\ref{sec:img_ps_result}, we first introduce the image cube, constructed with direct optimal mapping, and present the results for the window functions, 2D power spectrum, and 1D power spectrum.
We discuss future work in Section~\ref{sec:future_work} before concluding in Section~\ref{sec:conclusion}.
We use the \textit{WMAP} nine-year cosmology~\citep{bennett/etal:2013} throughout this paper.

\section{Power Spectrum Estimators} \label{sec:image_ps}
The key challenge for 21\,cm cosmology is to detect the EoR signals with the presence of foregrounds, which are at least four orders of magnitude brighter than the cosmological signals~\citep{liu/parsons/trott:2014a}.
Fortunately, the foreground and the EoR signals have different frequency properties within \textit{pixels}: foregrounds have smooth frequency spectra while the EoR signals have both smooth and fast-changing components.
One essential question is how well the EoR signal can be separated from smooth foregrounds.
The region, in Fourier space, free of foreground emission is called the \emph{EoR window}.

\subsection{Delay Power Spectrum} \label{subsec:delay_ps}
The technique of estimating the power spectrum by Fourier transforming the visibility's frequency axis is called delay power spectrum~\citep{parsons/etal:2012a}.
Visibilities are measured at different frequency channels, and instrument chromaticity introduces artifacts in the power spectrum.
For example, the main causes of instrument chromaticity are the frequency dependence of the primary antenna beam~\citep{sims/etal:2023} and the frequency dependence of the interferometric synthesized beam~\citep{Morales/etal:2012}.
These chromatic beams mix spatial structures into the frequency spectrum.
In particular, for the delay spectrum technique, the dependence of interferometric phase on frequency unavoidably introduces foreground power into the `wedge'~\citep{datta/bowman/carilli:2010, parsons/etal:2012a, parsons/etal:2012b, liu/parsons/trott:2014a, liu/parsons/trott:2014b}.

\subsection{Image Power Spectrum}

Alternatively, part of the instrumental chromaticity can be accounted for by analyzing frequency spectra of \emph{image pixels}, called image power spectrum~\citep{morales/hewitt:2004, liu/tegmark:2011, dillon/liu/tegmark:2013}. 
\cite{morales/etal:2019} compared delay and image power spectra and concluded that, assuming uniform $uv$ coverage, the image approach achieves lower foreground leakage into the wedge than the delay counterpart.

However, working with the pixel frequency spectrum requires forming images from visibilities.
Fundamentally, the imaging process presents the visibilities in a linear combination that minimizes the instrumental chromaticity.
How much instrumental chromaticity can be accounted for depends on the information we possess to reconstruct the sky.
For example, if we had perfect knowledge of the primary beam, we could remove the primary beam chromaticity in analysis; if we had complete $uv$-coverage, we could reconstruct each sky pixel without degeneracy.

There are available image power spectrum pipelines in the 21\,cm field.
We briefly review them before introducing our DOM image power spectrum.
We broadly consider pipelines that combine visibilities as image power spectrum pipelines, although some of them do not produce images.
In combining visibilities, the $uv$ gridding concept is the foundation of many image power spectrum pipelines~\citep{sullivan/etal:2012, offringa/etal:2014, trott/etal:2016, price:2024}.
Antenna arrays have fixed baseline physical lengths, thus $uv$ positions migrate with frequency.
Raw $uv$ sampling is often interpolated to a regular grid points for both imaging and power spectrum calculations.
Gridding averages measurements from nearby $uv$ positions, with the primary beam often used as the weight kernel.
Gridding estimates quantities that are not directly measured by the instrument.
Here are the three gridding-based image power spectrum pipelines:
\begin{enumerate}
    \item Fast Holographic Deconvolution (FHD)~\citep{sullivan/etal:2012} grids the measured visibilities into regularly-spaced $uv$ locations; it does not grid $w$-terms, so is limited to planar arrays.
    Then the regular $uv$ samples are Fast Fourier Transformed (FFT) into images for each frequency.
    The images are fed into the Error Propagated Power Spectrum with InterLeaved Observed Noise ($\epsilon ppsilon$)~\citep{barry/etal:2019} pipeline. 
    After averaging the images and converting the pixelization into HEALpix~\citep{gorski/etal:2005}, the images are Fourier transformed back to $uv$ space.
    At last, $\epsilon ppsilon$ uses Lomb-Scargle~\citep{lomb:1976, scargle:1982} periodogram to transform the line-of-sight axis from frequency into $k$ space, eventually to power spectra.
    In addition, \citet{dillon/etal:2015b} also uses the FHD images and calculates the power spectrum with a quadratic estimator.
    Covariance from the residual foreground is estimated from the images empirically; the quadratic estimator is then constructed to down-weight the residual foreground with the covariance matrix.
    \item Cosmology H\,I Power Spectrum (CHIPS)~\citep{trott/etal:2016} grids the raw $uv$ data including the curvature terms ($w$-terms).
    Without producing actual images, CHIPS uses least-square spectral analysis (LSSA) to transform the line-of-sight axis from frequency into $k$ space.
    The power spectrum is then estimated with a maximum likelihood (ML) estimator~\citep{trott/etal:2020}.\footnote{Interested readers are refered to \citet{jacobs/etal:2016} for a detailed comparison among CHIPS and the FHD-based power spectrum estimators.} 
    \item LOFAR~\citep{patil/etal:2017, mertens/etal:2020} uses W-Stacking Clean (\texttt{WSClean})~\citep{offringa/etal:2014} to convert gridded $uv$ data points to images.
    The \texttt{WSClean} is based on the iterative \texttt{CLEAN} algorithm~\citep{hogbom:1974, clark:1980, cornwell:2008, rau/cornwell:2011, mertens/etal:2020}, which focuses on estimating the fluxes of point sources.
    LOFAR's analysis uses Gaussian Process Regression (GPR)~\citep{mertens/Ghosh/Koopmans:2018} to correct the bias from the residual foreground.
    Finally, the 3D power spectrum is calculated from the 3D image cube with Fourier transform.
\end{enumerate}

DOM image power spectrum is fundamentally different from the existing methods in two ways: DOM maps visibilities at their original $uv$ locations without gridding~\citep{sullivan/etal:2012, barry/etal:2019, offringa/etal:2019};
the DOM mapping process is a simple linear operation without iteration~\citep{hogbom:1974, clark:1980, cornwell:2008, rau/cornwell:2011, mertens/etal:2020}.
\citet{murray/trott:2018} finds that dense and regular $uv$ coverage mitigates the wedge feature; however, the wedge reappears if the regular $uv$ locations are randomly offset by $10^{-5}$ of their lengths.
This counter-intuitive result likely comes from the gridding step, where the non-regular $uv$ sampling is estimated to a grid.
Our method calculates the images from their intrinsic $uv$ locations, avoiding potential induced errors from gridding.

The DOM algorithm involves only linear operations, robustly tracking the transfer function in the process; both the point spread function and the pixel covariance matrix can be accurately calculated.
One direct quantity to measure the foreground contamination in the EoR window is the power spectrum window function, which presents contributions from all the power spectrum space to one point.
The DOM algorithm provides the imaging transfer function which is essential for calculating the power spectrum window function; while
the other image power spectrum pipelines are not equipped to provide the power spectrum window functions.

\citet{gorce/etal:2022} calculated the window functions for the delay power spectrum; we will calculate the window functions for the DOM image power spectra in this paper.
The result quantitatively compares delay and image power spectrum regarding foreground contamination within the EoR window.
Furthermore, the robust calculation of the window functions enables us to normalize the image power spectrum without approximation.

\section{DOM Image Power Spectrum} \label{sec:dom_ps}

In this section, we introduce our power spectrum based on DOM.
We specifically focus on using the point-spread-function to calculate the power spectrum window function.

\subsection{Power Spectrum Definition} \label{subsec:ps_def}
We first present the power spectrum formula and discuss the symmetry assumptions for constructing a power spectrum and binning a 3D power spectrum into 2D and 1D.

With a 3D image cube, we calculate its 3D discrete Fourier transform (DFT) following the convention in \citet{mesinger/furlanetto:2007}:
\begin{equation} \label{equ:dft}
    \widetilde{m}(\mathbf{k}) = \frac{V}{N} \sum m(\mathbf{r}) e^{-i\mathbf{k \cdot r}},
\end{equation}
where $m(\mathbf{r})$ contains the intensity of the image cube and $\mathbf{r}$ is the 3D spatial vector, $\widetilde{m}(\mathbf{k})$ is the DFT of the image cube and $\mathbf{k}$ is the 3D wavenumber vector, $V$ is the physical size of the image cube (in the unit of Mpc$^3\cdot$h$^{-3}$), and $N$ is the number of voxels.

The power spectrum is calculated by squaring the DFT result and dividing out the total volume of the image cube~\citep{liu/shaw:2020}
\begin{equation} \label{equ:ps}
    P(\mathbf{k}) = \frac{\langle \widetilde{m}(\mathbf{k})^*\cdot \widetilde{m}(\mathbf{k}) \rangle}{V}.
\end{equation}
For spatially homogeneous signals, like cosmological signals from the early universe, all cross terms from $\langle \widetilde{m}(\mathbf{k'})^*\cdot \widetilde{m}(\mathbf{k}) \rangle$ average to zero, except when $\mathbf{k'} = \mathbf{k}$. 
Our observable universe is one random realization from its underlying power spectrum; the angle brackets indicate that the true power spectrum is the average over a large number of realizations.

The power spectrum is originally expressed in 3D $\mathbf{k}$ space. 
We bin the 3D power spectrum to 2D (or 1D) under different symmetry assumptions.
For the 2D binning, the two dimensions perpendicular to the line-of-sight (LOS) are circularly binned into $k_\bot$, assuming the two perpendicular dimensions share the same statistical properties.
Together with $k_\parallel$, parallel to LOS, we form the 2D $k_\bot-k_\parallel$ power spectrum, which is the natural space for foreground avoidance~\citep{datta/bowman/carilli:2010, Morales/etal:2012, parsons/etal:2012a}.
For the 1D binning, we assume isotropy of the signal, and all three dimensions are spherically binned into one dimension; the 3D vector $\mathbf{k}$ is condensed to only its length $k$.
Since the cosmological signals are believed to be homogeneous and isotropic at large scales, the 1D binning is often used to describe the power spectrum of the cosmological signals.
Homogeneity and isotropy in cosmological signals ensure the validity of the above steps; however, when the foregrounds are involved, which can be neither homogeneous nor isotropic, one should be aware of the underlining assumptions as we reduce power spectrum dimensions.

\subsection{Power Spectrum Estimator} \label{subsec:ps_estimator}
We form an estimated image cube with the direct optimal mapping algorithm.
Here we quickly review the method and refer to more details presented in \citetalias{xu/etal:2022}.
We first create the data model of the visibilities as
\begin{equation}
    \mathbf{d}_{f_i} = \mathbf{A}_{f_i} \mathbf{m}_{f_i} + \mathbf{n}_{f_i},
\end{equation}
where $\mathbf{d}_{f_i}$ is data, visibility, at the frequency channel $f_i$, $\mathbf{m}_{f_i}$ is the true sky emission at the frequency channel, $\mathbf{A}_{f_i}$ is the measurement matrix which describes how the interferometer integrates sky emission to produce visibilities, and $\mathbf{n}_{f_i}$ represents noise in the visibilities.

With this data model, the optimized sky recovery is described in a linear operation
\begin{equation}
    \mathbf{\hat{m}}_{f_i} = \mathbf{D}_{f_i} \mathbf{A}_{f_i}^\dagger \mathbf{N}_{f_i}^{-1} \mathbf{d}_{f_i},
\end{equation}
where $\mathbf{\hat{m}}_{f_i}$ is the estimated sky map, $\mathbf{D}_{f_i}$ is a normalization matrix, and $\mathbf{N}_{f_i}$ is the noise covariance matrix calculated as $\mathbf{N}_{f_i} = \langle \mathbf{n}_{f_i}^\dagger \mathbf{n}_{f_i} \rangle_n$.
The $\langle ... \rangle_n$ operator means that we are averaging over different visibility noise realizations.
Combining the above two equations, we define the point spread function (PSF) matrix $\mathbf{P}_{f_i}$
\begin{eqnarray}
    \langle \mathbf{\hat{m}}_{f_i} \rangle_n &=& \mathbf{D}_{f_i} \mathbf{A}_{f_i}^\dagger \mathbf{N}_{f_i}^{-1} (\mathbf{A}_{f_i} \mathbf{m}_{f_i} + \langle \mathbf{n}_{f_i} \rangle_n) \nonumber \\
    &=& ( \mathbf{D}_{f_i} \mathbf{A}_{f_i}^\dagger \mathbf{N}_{f_i}^{-1} \mathbf{A}_{f_i})\, \mathbf{m}_{f_i} \nonumber \\
    &\equiv& \mathbf{P}_{f_i}\, \mathbf{m}_{f_i},
\end{eqnarray}
where we use the property that noise averages to zero ($\langle \mathbf{n}_{f_i} \rangle_n = 0$), and the last line defines the PSF matrix as $\mathbf{P}_{f_i} = \mathbf{D}_{f_i} \mathbf{A}_{f_i}^\dagger \mathbf{N}_{f_i}^{-1} \mathbf{A}_{f_i}$.
The above formalism was introduced in \citetalias{xu/etal:2022}; below, we introduce the power spectrum estimator.

The map pixels are chosen to be smaller than the synthesized beam to ensure that the measurement resolution is not limited by the mapping.
The subscript $f_i$ in all the variables shows that they are for one frequency channel.
Sky images are generated for each frequency channel, and the frequency channels are converted into LOS distances.
Combining the 2D sky images at different LOS distances yields a 3D image cube --- $\mathbf{\hat{m}}$.
It is related to the true 3D image cube $\mathbf{m}$ by the 3D PSF matrix $\mathbf{P}$.
We aggregate the 2D image PSF matrix $\mathbf{P}_{f_i}$ to form the 3D PSF matrix with $\mathbf{P}_{f_i}$ as block-diagonal matrices
\begin{equation}
    \mathbf{P} = \mathrm{diag}(\mathbf{P}_{f_1}, ..., \mathbf{P}_{f_i}, ..., \mathbf{P}_{f_N}).
\end{equation}
Here we do not consider frequency-frequency correlations.
The aggregated $\mathbf{P}$ maps the true 3D image cube to the estimation
\begin{equation} \label{equ:dirty_and_sky}
    \langle \mathbf{\hat{m}} \rangle_n = \mathbf{P m},
\end{equation}
where the 3D image cube $\mathbf{m}$ and its estimation $\mathbf{\hat{m}}$ are flattened to column vectors. 

The image cube estimation is then tapered along the three dimensions.
The tapering apodizes hard boundaries, suppressing ringing during Fourier transformation.
Meanwhile, we treat all the power within each voxel as a single point source~\citep{liu/tegmark:2011, dillon/liu/tegmark:2013}, and we will correct for the voxel window function later.
With the Fourier transform convention in Equation~\ref{equ:dft} and \ref{equ:ps}, we define the quadratic estimator of the image cube to be
\begin{equation} \label{equ:quadratic_estimator}
    \hat{q}_\alpha = \frac{V}{N^2} \mathbf{\hat{m}}^\dagger \mathbf{R}^\dagger \mathbf{E}_\alpha \mathbf{R} \mathbf{\hat{m}},
\end{equation}
where $\mathbf{R}$ is the 3D tapering function of the image cube.
Then, we define 
\begin{equation}
    \mathbf{E}_\alpha = \mathbf{c}^\dagger_\alpha \mathbf{c}_\alpha,
\end{equation}
where $\mathbf{c}_\alpha$ is the 3D discrete Fourier transform operator at $\mathbf{k}_\alpha$
\begin{equation}
    \mathbf{c}_\alpha = (e^{-i \mathbf{k}_\alpha \cdot \mathbf{r}_1}, ..., e^{-i \mathbf{k}_\alpha \cdot \mathbf{r}_j}, ..., e^{-i \mathbf{k}_\alpha \cdot \mathbf{r}_{N}}).
\end{equation}
Elements within $\mathbf{c}_\alpha$ cover all voxels $\mathbf{r}_i$ within the image cube; multiplying $\mathbf{c}_\alpha$ on an image cube vector gives the Fourier transform of the image cube at $\mathbf{k}_\alpha$.
On the contrary, the conjugate transpose operator $\mathbf{c}_\alpha^\dagger$ is the inverse 3D discrete Fourier transform
\begin{equation}
    \mathbf{c}_\alpha^\dagger = (e^{i \mathbf{k}_\alpha \cdot \mathbf{r}_1}, ..., e^{i \mathbf{k}_\alpha \cdot \mathbf{r}_j}, ..., e^{i \mathbf{k}_\alpha \cdot \mathbf{r}_{N}})^T.
\end{equation}
Neither $\mathbf{c}_\alpha$ nor $\mathbf{c}_\alpha^\dagger$ contains DFT normalization factors; instead, the normalization factors are explicitly represented in the leading factors of Equation~\ref{equ:quadratic_estimator}.

Coming back to the quadratic estimator $\hat{q}_\alpha$, plugging in the $\mathbf{m}$ and $\mathbf{\hat{m}}$ relation of Equation~\ref{equ:dirty_and_sky}, the quadratic estimator is written as
\begin{equation} \label{equ:quadratic_true}
    \langle \hat{q}_\alpha \rangle_n = \frac{V}{N^2} \mathbf{m}^\dagger \mathbf{P}^\dagger \mathbf{R}^\dagger \mathbf{E}_\alpha \mathbf{R} \mathbf{P} \mathbf{m}.
\end{equation}
With the true sky brightness, we define the voxel covariance $\mathbf{C \equiv \langle m\, \mathbf{m}^\dagger \rangle}$.
The $\langle...\rangle$ operator indicates we are averaging over different sky realization, as initially defined in Equation~\ref{equ:ps}.
This covariance matrix is associated with the power spectrum, and can be expressed as a linear combination of the band powers~\citep{dillon/liu/tegmark:2013, liu/shaw:2020}
\begin{equation}
    \mathbf{C \equiv \langle m\, \mathbf{m}^\dagger \rangle} = \frac{1}{V} \mathbf{\sum_\beta} p_\beta \mathbf{E_\beta} \cdot |\Phi(\mathbf{k_\beta})|^2,
\end{equation}
The above equation adds band powers across the $\mathbf{k}$-space, and the normalization factor $1/V$ is the $\mathbf{k}$-space resolution for the summation.
In addition, the $\Phi(\mathbf{k_\beta})$ function is the voxel window function at the $\mathbf{k}_\beta$ position.
We use the 3D boxcar profile as the real-space voxel profile, therefore, $\Phi(\mathbf{k_\beta})$ is expressed as the product of three sinc functions:
\begin{equation}
    \Phi(\mathbf{k}) \equiv j_0\left( \frac{k_x \Delta x}{2} \right) j_0\left( \frac{k_y \Delta y}{2} \right) j_0\left( \frac{k_z \Delta z}{2} \right),
\end{equation}
where $j_0(x) = \mathrm{sin}(x)/x$; $k_x, k_y, k_z$ are the three components of the $\mathbf{k}$ vector, and $\Delta x, \Delta y, \Delta z$ are the real-space resolution of the image cube for each dimension~\citep{dillon/liu/tegmark:2013}.

Equation~\ref{equ:quadratic_true} has scalars on both sides, so we add the trace operator on both sides.
With the cyclic property of the trace, we can circularly shift the matrices on the right-hand side.
Also using the above equation, we rewrite Equation~\ref{equ:quadratic_true}
\begin{eqnarray}
    \langle \hat{q}_\alpha \rangle &=& \frac{V}{N^2} \mathrm{tr} [\langle \mathbf{m}^\dagger \mathbf{P}^\dagger \mathbf{R}^\dagger \mathbf{E}_\alpha \mathbf{R} \mathbf{P} \mathbf{m} \rangle] \nonumber \\
    &=& \frac{V}{N^2} \mathrm{tr} [\mathbf{P}^\dagger \mathbf{R}^\dagger \mathbf{E}_\alpha \mathbf{R} \mathbf{P} \langle \mathbf{m} \mathbf{m}^\dagger \rangle] \nonumber \\
    &=& \frac{V}{N^2} \mathrm{tr} [\mathbf{P}^\dagger \mathbf{R}^\dagger \mathbf{E}_\alpha \mathbf{R} \mathbf{P} \frac{1}{V} \mathbf{\sum_\beta} p_\beta \mathbf{E_\beta} |\Phi(\mathbf{k_\beta})|^2 ] \nonumber \\
    &=& \frac{1}{N^2} \mathbf{\sum_\beta} \mathrm{tr} [\mathbf{P}^\dagger \mathbf{R}^\dagger \mathbf{E}_\alpha \mathbf{R} \mathbf{P} \mathbf{E_\beta} |\Phi(\mathbf{k_\beta})|^2] p_\beta \nonumber \\
    &\equiv& \sum_\beta H_{\alpha \beta} p_\beta .
\end{eqnarray}
Please note that we are averaging over both noise realizations and sky realizations above.
Here we define the power spectrum response function $\mathbf{H}$, which transfer the true sky power to our quadratic estimator.
Element-wise, $\mathbf{H}$ is written as
\begin{eqnarray}
    H_{\alpha \beta} &=& \frac{|\Phi(\mathbf{k_\beta})|^2}{N^2} \mathrm{tr}(\mathbf{P^\dagger R^\dagger E_\alpha R P E_\beta} ) \nonumber \\
    &=& \frac{|\Phi(\mathbf{k_\beta})|^2}{N^2} \mathrm{tr} (\mathbf{P^\dagger R^\dagger c_\alpha^\dagger c_\alpha R P c_\beta^\dagger c_\beta}).
\end{eqnarray}
Using the matrix trace cyclic property again, we can rewrite the above equation
\\
\\
\begin{eqnarray} \label{equ:reponse_matrix}
    H_{\alpha \beta} &=& \frac{|\Phi(\mathbf{k_\beta})|^2}{N^2} \mathrm{tr} (\mathbf{c_\beta P^\dagger R^\dagger c_\alpha^\dagger c_\alpha R P c_\beta^\dagger}) \nonumber \\
    &=& \frac{|\Phi(\mathbf{k_\beta})|^2}{N^2} \mathrm{tr}[(\mathbf{c_\alpha R P c_\beta^\dagger)^\dagger (c_\alpha R P c_\beta^\dagger})] \nonumber \\
    &=& |\Phi(\mathbf{k_\beta})|^2 \, \mathrm{tr}[(\frac{\mathbf{c_\alpha}}{\sqrt{N}} \mathbf{R P} \frac{\mathbf{c_\beta^\dagger}}{\sqrt{N}})^\dagger (\frac{\mathbf{c_\alpha}}{\sqrt{N}} \mathbf{R P} \frac{\mathbf{c_\beta^\dagger}}{\sqrt{N}})] \nonumber \\
    &=& |\Phi(\mathbf{k_\beta})|^2 \, |\frac{\mathbf{c_\alpha}}{\sqrt{N}} \mathbf{R P} \frac{\mathbf{c_\beta^\dagger}}{\sqrt{N}}|^2,
\end{eqnarray}
The last line indicates the normalization for DFT: we divide forward and background DFT by $\sqrt{N}$. 
Equation~\ref{equ:reponse_matrix} is the formula to calculate the response matrix $\mathbf{H}$.

The $\mathbf{H}$ matrix links the quadratic estimator $\mathbf{\hat{q}}$ and the true power spectrum $\mathbf{p}$ as
\begin{equation} \label{equ:qe_to_ps}
    \mathbf{\langle \hat{q} \rangle = H p}.
\end{equation}
The quadratic estimator $\mathbf{\hat{q}}$ is an intermediate product of the power spectrum estimation, and we need the response matrix $\mathbf{H}$ to convert it to the power spectrum estimator $\mathbf{\hat{p}}$.
Equation~\ref{equ:reponse_matrix} also shows that $\mathbf{H}$ is determined by $\mathbf{RP}$ --- the transfer function in image space.
Essentially, $\mathbf{H}$ and $\mathbf{RP}$ represent the same relation in two domains.

\subsection{Window Functions} \label{subsec:window_function}
Window functions describe, in power spectrum space, how the true sky power is mapped into the measured power spectrum.
Ideally, window functions are delta functions, which is not always achievable.
In practice, one band power takes contribution from all power spectrum parameter space.
The contribution is characterized by a set of weights, which is the window function.

Continuing with Equation~\ref{equ:qe_to_ps}, if we could construct a matrix $\mathbf{M}$ to invert $\mathbf{H}$, unambiguously recovering the underlining band power $\mathbf{p}$ from $\mathbf{\hat{q}}$, we would equivalently obtain delta-function-like window functions.
In practice, $\mathbf{H}$ is generally not invertible for interferometers.
Therefore, different forms of $\mathbf{M}$ are constructed for various purposes~\citep{seljak:1998, tegmark/hamilton/xu:2002, liu/tegmark:2011, ali/etal:2015, hera/etal:2021, kern/liu:2021}.
Here we sum the columns for each row in the $\mathbf{H}$ matrix and divide out the sum for each element in the corresponding row --- essentially normalizing the integrated power.
The detailed contribution in the $\mathbf{H}$ matrix allows us to precisely calculate the normalization factor for each power band.
Mathematically, we form a diagonal matrix $\mathbf{M}$ with each element as the inverse of the row sum\footnote{Analogous to normalizing the weights for a weighted sum.}
\begin{equation} \label{equ:m_mat}
    \mathbf{M} = \mathrm{diag}(1/\sum_\beta H_{1, \beta},...,1/\sum_\beta H_{i, \beta},...,1/\sum_\beta H_{N, \beta}).
\end{equation}
The full-rank $\mathbf{M}$ matrix takes our quadratic estimator $\mathbf{ \hat{q} }$ to the estimated band powers $\mathbf{\hat{p}}$~\citep{tegmark:1997, liu/parsons/trott:2014a}
\begin{equation} \label{equ:ps_estimator}
    \mathbf{\hat{p} = M \hat{q}}.
\end{equation}
With this definition of $\mathbf{M}$, the window function matrix is defined as $\mathbf{W = M H}$, the normalized power is expressed as
\begin{equation}
    \mathbf{\langle \hat{p} \rangle = M \langle \hat{q} \rangle =  M H p} = \mathbf{W p}.
\end{equation}

The window function matrix $\mathbf{W}$ is fully defined by $\mathbf{H}$
\begin{eqnarray} \label{equ:window_function}
    W_{\alpha \beta} &=& \sum_\gamma M_{\alpha \gamma} H_{\gamma \beta} \nonumber \\
                     &=& M_{\alpha \alpha} H_{\alpha \beta} \nonumber \\
                     &=& \frac{H_{\alpha \beta}}{\sum_{\delta} H_{\alpha \delta}}.
\end{eqnarray}
The above equation also shows that weights in each row of $\mathbf{W}$ sums to one.

Window functions in the delay power spectrum were studied in
\citet{parsons/etal:2012b, liu/parsons/trott:2014a, gorce/etal:2022}; here, we investigate window functions from this image power spectrum.

\section{Image Power Spectrum Application} \label{sec:img_ps_result}
Equipped with the above formalism, we investigate the properties of the image power spectrum.
We map simulated noiseless visibilities to one image cube and calculate its image power spectrum along with the window functions.
We compare the window functions with the delay spectrum ones to understand the difference between the two estimators.
We show the 2D/1D power spectrum to evaluate the foreground separation.

\subsection{The Image Cube} \label{subsec:img_cube}
We use the HERA validation simulations~\citep{aguirre/etal:2021}, which is based on HERA Phase I array configuration with 33 unflagged antennas~\citep{hera/etal:2021, hera/etal:2022}.
We choose three datasets with different sky models, including \textit{EoR-only}, \textit{foregrounds-only}, and \textit{EoR+foregrounds}.
Please note that the EoR signals in these simulations are boosted compared to the fiducial model~\citep{aguirre/etal:2021}. 
The simulation includes neither noise nor systematics.
We select the data from 180 time-integration when the zenith centers around 24.2\dg{} in R.A.; and we select a 17.5\,MHz frequency range from 150.34\,MHz to 167.84\,MHz (with 0.098\,MHz resolution). 
After Blackman-Harris tapering, the effective bandpass is around 8\,MHz, within which cosmological evolution is insignificant.
The frequency range and resolution are consistent with Band\,2 defined in \citet{hera/etal:2021}. 
The data contain both East-West and North-South polarization, we only analyze the East-West polarization in this paper.

We use the Fast Fourier Transform (FFT) algorithm to calculate the DFT, which requires a regular grid. 
Therefore, we create a regular R.A./Decl. grid in the sky-plane.
However, the pixels do not have equal areas on a curved surface; their solid angle changes proportionally to the cosine of the pixel declination.

\begin{figure}
    \centering
    \includegraphics[width=\linewidth]{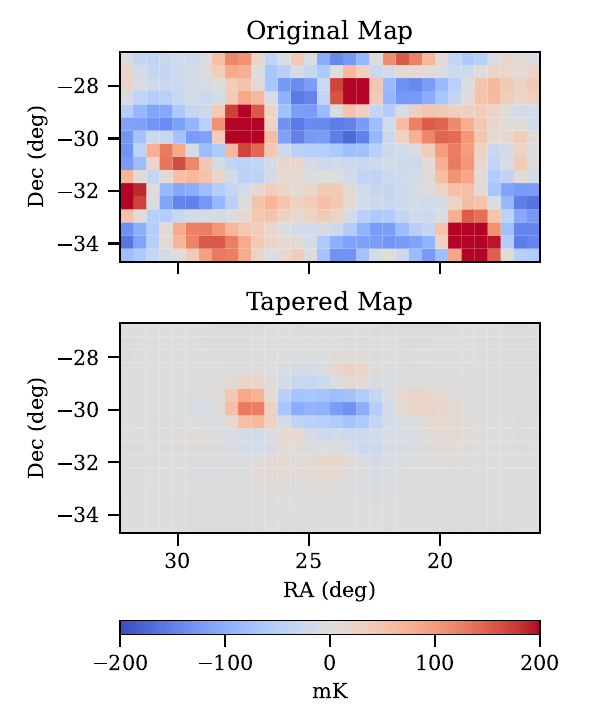}
    \caption{Original and tapered maps at 159.04\,MHz.
    We use the direct optimal mapping algorithm~\citepalias{xu/etal:2022} to map the noiseless simulation data to regular R.A./Decl. grids.
    The original map consists of $32\times16=512$ pixels, covering $16^\circ \times 8^\circ = 128$ square degrees.
    After tapering with the Blackman-Harris function along three dimensions, the off-center signals are highly attenuated to suppress ringing structures from Fourier transform.
    We select the center frequency for plotting where the frequency tapering effect is minimal.
    In total, we map 180 maps in the frequency range from 150.34\,MHz to 167.84\,MHz.
    }
    \label{fig:map}
\end{figure}

For each frequency channel, the visibilities are mapped with the direct optimal mapping algorithm.
The grid is chosen to center at 24.2\dg{} in R.A. and -30.7\dg{} in Decl., with 0.5\dg{} resolution along both coordinate directions (although the physical distance is less than 0.5\dg{} along the R.A. direction).
The 0.5\dg{} resolution is chosen to be half of the size of the $\sim$1\dg{} synthesized beam.
The final map contains $32\times16=512$ pixels, covering $16^\circ \times 8^\circ = 128$ square degrees.
The top panel of Figure~\ref{fig:map} shows the map at the central frequency channel --- 159.04\,MHz.
The mapping repeats for the 180 frequency channels.

The frequency channels are converted to redshifts and to comoving distances in the unit of $\mathrm{Mpc \cdot h^{-1}}$.
Uniform sampling in frequency leads to nonuniform sampling in comoving distances, because of the nonlinear relation between frequency and comoving distance.
The distance intervals differ up to $\sim$5.5\% across the whole range, we chose the average as the resolution along the LOS direction.
In addition, comoving distances determine the physical spacing in the sky-plane.
Within our frequency range, the comoving distance changes up to $\sim$3.2\%; we use the average LOS comoving distance to calculate the sky-plane resolution.
Finally, the voxel size is calculated by multiplying the physical resolutions along three dimensions, which is $\sim3.5\times10^3\, \mathrm{Mpc^3 \cdot h^{-3}}$.
Table~\ref{tab:img_cube} collects the related parameters of the image cube.
Please note the difference in comoving resolutions between the LOS and the sky-plane: the frequency channels provide much finer comoving resolution compared to the angular solution.

\begin{deluxetable}{lccc}
\tablecaption{Parameters for the Image Cube\label{tab:img_cube}}
\tablehead{
\colhead{Parameters} & 
\colhead{LOS} & 
\colhead{R.A.} &
\colhead{Decl.}
}
\startdata
$N_\mathrm{pixel}$ & 180 & 32 & 16 \\
Range & 17.5\,MHz & 16\dg{} & 8\dg{} \\
Resolution & 0.098\,MHz & 0.5\dg{} & 0.5\dg{} \\
Range ($\mathrm{Mpc \cdot h^{-1}}$) & 205.4 & 1525 & 886 \\
Resolution ($\mathrm{Mpc \cdot h^{-1}}$) & 1.15 & 47.7 & 55.4 \\
\enddata
\tablecomments{Image cube parameters shown in measurement units, including angular/frequency resolution, and the related comoving distances.
We use the average comoving distance to calculate sky-plane resolution; we use the average LOS interval as the LOS resolution.}
\vspace{-1cm}
\end{deluxetable}

\subsection{Window Function Results} \label{subsec:window_function_calc}

\begin{figure*}
    \centering
    \includegraphics[width=0.8\linewidth]{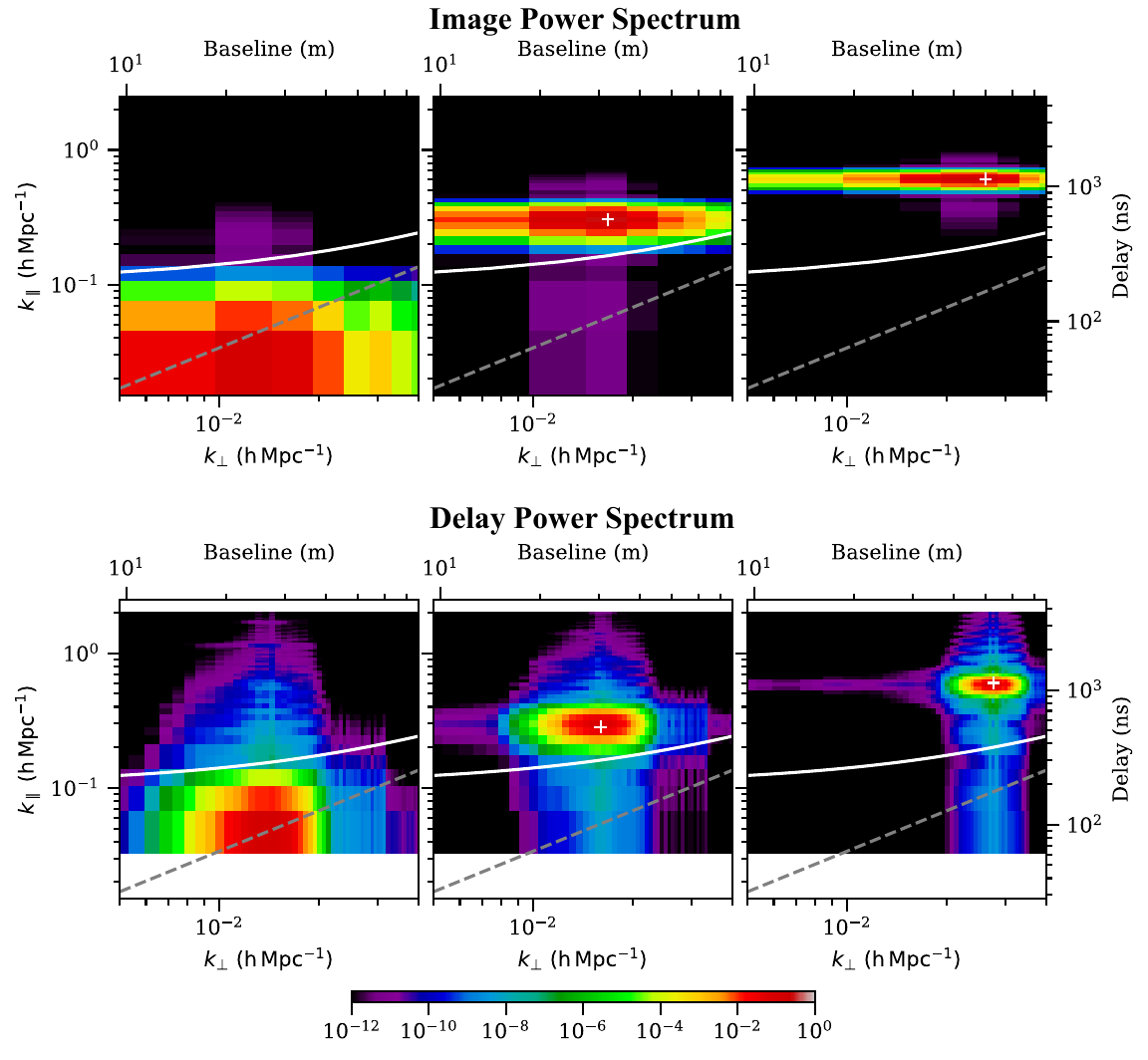}
    \caption{Window functions for image power spectrum and delay power spectrum.
    The top panels show image window functions at three $(k_\bot, k_\parallel)$ points; the bottom panels show delay window functions at similar $(k_\bot, k_\parallel)$ locations~\citep{gorce/etal:2022}.
    Locations of the native $(k_\bot, k_\parallel)$ bins are marked with white plus signs (the symbols are out of the plot range in the first column).
    All panels share the same logarithmic scales in two axes, as well as the color range from 1 to $10^{-12}$.
    The dashed gray lines shows the horizon wedge and the solid white lines show 200\,ns beyond the wedge.
    Both sets of window functions are based on the same array configuration and observation parameters.
    The $k_\bot - k_\parallel$ binning is different from the image and delay power spectrum; white space is displayed where delay window functions do not have data.
    All the window functions are sum-normalized.
    The top and right axes show the corresponding baseline and delay values for $k_\bot$ and $k_\parallel$ respectively.
    The image window functions show lower power leakage along the $k_\parallel$ direction ($<10^{-11}$) compared to the delay window functions, which is critical for foreground avoidance; the spread along the $k_\bot$ direction is from the limited sky patch of the image cube.
    }
    \label{fig:window_function_3pts}
\end{figure*}

Now we look at the image power spectrum window functions for the HERA Phase I configuration.
Section~\ref{subsec:window_function} gives the theoretical formalism of the window function; here we present the actual calculation.

The aggregated $\mathbf{P}$ matrix is a square matrix with $(N_\mathrm{pixel} \times N_\mathrm{freq.})^2 = (512 \times 180)^2 \approx 8\times10^9$ elements.
Storing and operating this matrix is the most computationally expensive step in our image power spectrum analysis.
The tapering matrix $\mathbf{R}$ applies tapering functions along the three dimensions of the original image cube.
Here we use the Blackman-Harris (4-term) tapering function along the three dimensions.
A tapered map, perpendicular to LOS, is shown in the bottom panel of Figure~\ref{fig:map}.

To calculate $\mathbf{H}$, we pick individual columns in the $\mathbf{RP}$ matrix, which are 3D vectors flattened to 1D.
We reshape the 1D columns back into 3D and calculate their 3D FFT.
The result is then flattened to replace the same column.
The calculation is repeated for all columns of $\mathbf{RP}$.
Correspondingly, inverse FFT is performed across all rows from the previous result.
Then, we square the individual elements of the matrix.
Finally, we calculate the pixel window function $|\Phi(\mathbf{k})|^2$ and apply the result to all rows of the previous matrix to obtain $\mathbf{H}$.

Summing up columns in $\mathbf{H}$, we calculate the $\mathbf{M}$ matrix according to Equation~\ref{equ:m_mat}.
The $\mathbf{M}$ matrix normalizes the quadratic estimator to get the power spectrum estimator as in Equation~\ref{equ:ps_estimator}.
Finally, the window function matrix $\mathbf{W}$ is calculated as $\mathbf{W = MH}$.
\begin{figure*}
    \centering
    \includegraphics[width=\linewidth]{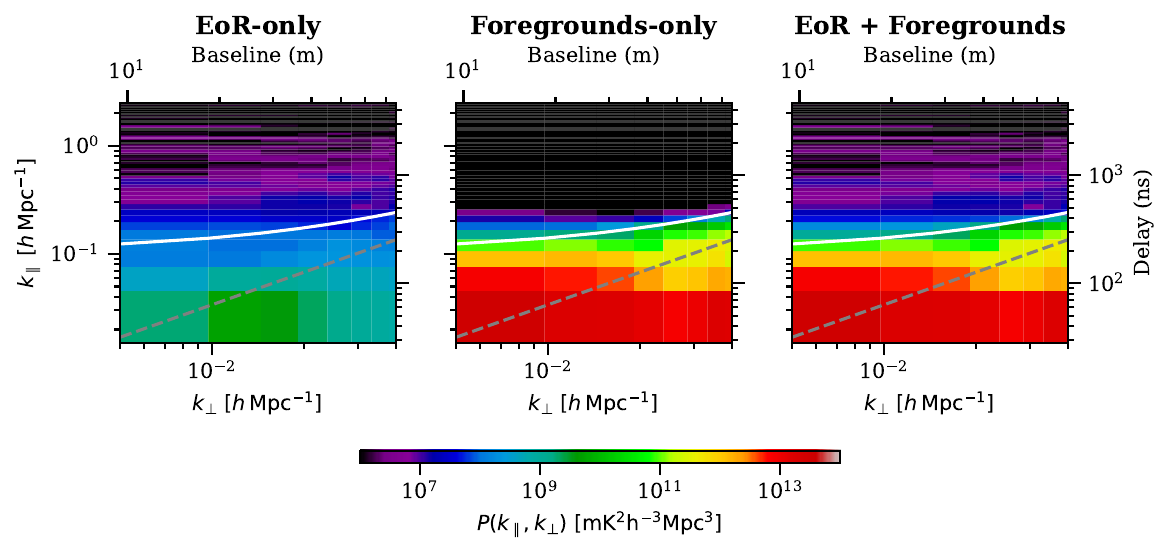}
    \caption{2D image power spectra from EoR-only, foregrounds-only, and EoR+foregrounds simulations.
    The three power spectra were calculated from simulated visibilities.
    The power spectra share the same $k_\bot$, $k_\parallel$ range as well as the color range; the dashed line shows the horizon wedge and the solid line shows 200\,ns beyond the wedge.
    The EoR power spectrum shows power distributed across the whole space, with more power at low-$k_\parallel$. 
    The foregrounds' power is concentrated within low-$k_\parallel$ region, with a window not contaminated in high-$k_\bot$ regions.
    Finally, the EoR+foregrounds power spectrum shows the result with both sky components: foreground features dominate low-$k_\parallel$ and EoR features prevail in the high-$k_\parallel$ EoR window.
    }
    \label{fig:normalized_2d_ps}
\end{figure*}
The window function matrix $\mathbf{W}$ is defined in 3D band powers; for each band power at a certain $\mathbf{k}$ bin, its window function shows the contribution from all $\mathbf{k}$ bins whose contribution sums to unity.
The 3D window functions are then reduced to 2D with symmetry.
The two sky-plane dimensions are circularly binned into $k_\bot$; the LOS direction is kept unchanged as $k_\parallel$.
Values from related 3D $\mathbf{k}$ bins are summed, because the values record contribution instead of intensity, into ($k_\bot$, $k_\parallel$) space with equal weights.
The $k_\parallel$ direction is divided into 90 linear bins from 0 to 2.71\,$\mathrm{h}\cdot\mathrm{Mpc}^{-1}$; the $k_\bot$ direction is divided into 16 linear bins from $7.4\times10^{-3}$ to $7.8\times10^{-2}\,\mathrm{h}\cdot\mathrm{Mpc}^{-1}$.

The top panels of Figure~\ref{fig:window_function_3pts} show image power spectrum window functions at three ($k_\bot, k_\parallel$) points.
Along the $k_\bot$ direction, the window functions show wide spreading.
This is because we currently only consider a small sky patch as shown in Figure~\ref{fig:map}.
The small sky patch and the sky-plane tapering functions generate the extended $k_\bot$ kernels.

Along the $k_\parallel$ direction, contributions concentrate around the local power band; distant power bands contribute $<10^{-11}$ of unity.
This is critical for separating EoR from foregrounds: the smooth foregrounds only occupy low-$k_\parallel$ regions while the EoR signal takes up the entire $k_\bot-k_\parallel$ space;
therefore, the detectability of the EoR signal lies in the assumption that power measured within the EoR window has minimal contribution from low-$k_\parallel$ regions.
The window functions in Figure~\ref{fig:window_function_3pts} directly show how much power leaks from low-$k_\parallel$ to high-$k_\parallel$.

The bottom panels of Figure~\ref{fig:window_function_3pts} show window functions of the same datasets at similar ($k_\bot, k_\parallel$) locations from the delay power spectrum~\citep{gorce/etal:2022}.
The window functions are also normalized by integrated power; the same Blackman-Harris tapering is applied along the visibility frequency axis.
The distant $k_\parallel$ contribution is approximately three orders of magnitude higher than the image power spectrum.
However, along the $k_\bot$ direction, the window functions display a more compact kernel compared to the image power spectrum.
This is because visibilities integrate the whole sky and the delay spectrum analyzes visibilities directly, equivalently measuring the full sky weighted by the primary beam.

\subsection{2D Power Spectrum} \label{subsec:2d_ps}
The previous section examines the window functions from three representative $(k_\bot, k_\parallel)$ bins,
now we investigate the full 2D image power spectra.

We first calculate the quadratic estimator $\mathbf{\hat{q}}$ with Equation~\ref{equ:quadratic_estimator}.
We FFT and inverse FFT the tapered image cube and multiply the results together element-wise, then we multiply the $V/N$ factor\footnote{The other $N$ in the denominator is included in the FFT and iFFT normalization.} to obtain the 3D quadratic estimator $\mathbf{\hat{q}}$.
We use the matrix $\mathbf{M}$ to normalize the quadratic estimator for the power spectrum estimator (Equation~\ref{equ:ps_estimator}).
After obtaining the normalized 3D power spectrum estimator, we bin it to the $(k_\bot, k_\parallel)$ space by averaging the related 3D band power.
In the averaging step, we use equal weights.
The power spectrum in the 2D $(k_\bot, k_\parallel)$ space are shown in Figure~\ref{fig:normalized_2d_ps}.

The EoR-only power spectrum shows power across the entire $k_\bot - k_\parallel$ space.
Not much change is observed along the $k_\bot$ direction while there is a clear decreasing trend from low-$k_\parallel$ to high-$k_\parallel$.
The peak of the 2D power spectrum is around $10^{10} \mathrm{mK^2\cdot h^{-3} Mpc^3}$, which is amplified above the fiducial theoretical model in the simulation~\citep{aguirre/etal:2021}.
The foregrounds-only power spectrum shows a different situation: the power is constrained within the low-$k_\parallel$ regions, mostly within 200\,ns beyond the horizon wedge.
The peak of the foregrounds-only power spectrum rises as high as $10^{14}\, \mathrm{mK^2\cdot h^{-3} Mpc^3}$, four orders of magnitude higher than the simulated EoR-only peak.
However, its power does not leak into the high-$k_\parallel$ region with $>10^8$ dynamic range.
This is consistent with the window function result above:
the window functions show $<10^{-11}$ leakage from low-$k_\parallel$ to high-$k_\parallel$, which ensures the overwhelming power from the smooth foregrounds does not affect the EoR window.

Finally, we come to the power spectrum of EoR+foregrounds.
Please note that this power spectrum is not obtained by simply adding the previous two power spectra; instead, we start from visibilities, form the image cubes, and calculate the 2D power spectrum.
In this power spectrum, the low-$k_\parallel$ region is dominated by the foregrounds, resembling the features in the foregrounds-only power spectrum; the high-$k_\parallel$ region is filled by EoR signals.
The clear separation between the foregrounds and the EoR further illustrates the detectability of the EoR with the image power spectrum.

\begin{figure}
    \centering
    \includegraphics[width=\linewidth]{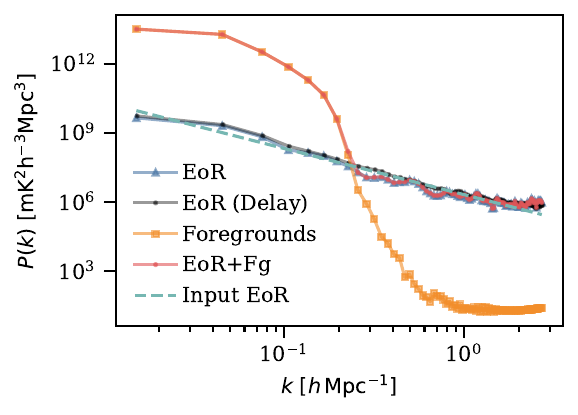}
    \caption{1D power spectrum measured from three simulation datasets.
    The foreground signal dominates in the low-$k$ regions; the boosted EoR signal dominates in the high-$k$ region.
    The foreground-only power spectrum shows the $10^{12}$ dynamic range from low-$k$ to high-$k$.
    Also shown is the EoR-only spectrum measured by the delay power spectrum.
    Both the image power spectrum and the delay power spectrum agree with the input power, represented in a dashed line.
    The deviation of the measurement at high-$k$ is due to aliasing from DFT~\citep{aguirre/etal:2021}.
    The deviation from a straight line in the mid-$k$ range is due to the small sky patch selected for the image power spectrum.
    }
    \label{fig:1dps}
\end{figure}

\subsection{1D Power Spectrum} \label{subsec:1d_ps}
The EoR power spectrum is often collapsed into 1D assuming it is homogeneous and isotropic.\footnote{Please refer to Section~\ref{subsec:ps_def} for discussions on symmetry and power spectrum dimension reduction.}
We present the 1D power spectrum in this section by binning the original 3D power spectrum.
The binning is based on equal-weight averaging, similar to the 2D power spectrum.

Figure~\ref{fig:1dps} shows the 1D power spectrum from EoR-only, foregrounds-only, and EoR+foregrounds simulations.
The EoR+foregrounds 1D power spectrum closely follows the foregrounds-only spectrum in low-$k$ and the EoR-only spectrum in high-$k$.
In high-$k$, the EoR signal is more than five orders of magnitude stronger than that of the foregrounds.
The foreground avoidance capability of the image power spectrum is illustrated in this figure.
The foreground-only spectrum shows the $10^{12}$ dynamic range achievable from this estimator: at low-$k$ the foregrounds are at $>10^{13}\, \mathrm{mK}^2\cdot\mathrm{h}^{-3}\cdot\mathrm{Mpc}^3$ and decrease to $10^{1}\,\mathrm{mK}^2\cdot\mathrm{h}^{-3}\cdot\mathrm{Mpc}^3$ at high-$k$.
The smooth foregrounds are suppressed by twelve orders of magnitude.
Also plotted is the 1D EoR-only power spectrum measured with the delay power spectrum, and the input EoR power.
The image power spectrum and delay power spectrum both recover the input EoR power.

\begin{figure}
    \centering
    \includegraphics[width=\linewidth]{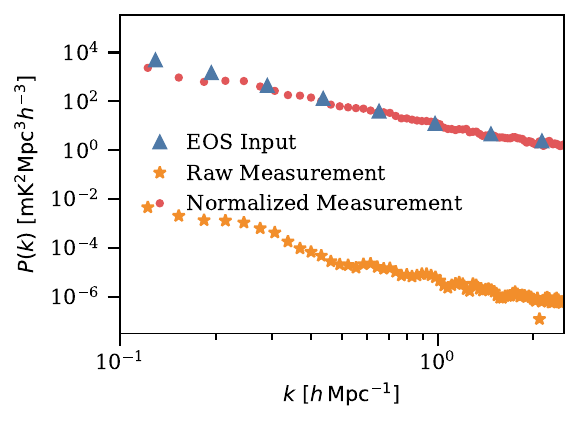}
    \caption{1D Power Spectrum from the evolution of 21\,cm structure (EOS) simulation.
    We downsample and convolve the EOS image cube and measure the 1D image power spectrum.
    The normalized 1D power spectrum agrees with the input after being adjusted by six orders of magnitude from the normalization step.
    }
    \label{fig:eos_1dps}
\end{figure}

In addition, we use the Evolution Of 21\,cm Structure (EOS)\footnote{URL: \url{http://homepage.sns.it/mesinger/EOS.html}} simulation~\citep{sobacchi/mesinger:2014} to further validate the image power spectrum 1D power spectrum.
The EOS project provides image cubes and also their input 1D power spectrum.
Therefore, we can calculate the 1D image power spectrum and compare it with the input.
The EOS simulation provides image cubes with 1.6\,Gpc (1.11\,$\mathrm{Gpc \cdot h^{-1}}$) and 1024 bins on each side.
We use one image cube at redshift 5.76. 
Along the x-axis, we downsample the EOS image cube into 32 bins; along the y-axis, we also downsample the EOS image cube into 32 bins but select the middle 16 bins; along the z-axis, we select the first 180 bins without downsampling.
We average the intensity among associated voxels during downsampling.
This gives us the $180\times32\times16$ image cube similar to the DOM image cube.
Then we convolve the downsampled EOS image cube with the $\mathbf{P}$ matrix to get the instrument-convolved image cube.

We measure the image power spectrum and bin the 3D power spectrum into 1D.
Figure~\ref{fig:eos_1dps} shows the input 1D power spectrum and the normalized 1D power spectrum from DOM.
The measurement recovers the input 1D power spectrum across the majority of the $k$, with some mismatch at the low-$k$ end.
The inconsistency at the low-$k$ end comes from the edge effect of the finite EOS image cube.
Also shown in Figure~\ref{fig:eos_1dps} is the 1D power spectrum before normalization.
The comparison shows the normalization correctly rescales the power spectrum over six orders of magnitude.

\section{Future Work} \label{sec:future_work}
We have developed the image power spectrum pipeline and understand its window function and power spectra with noiseless simulations.
We will investigate the effects of realistic noise, radio-frequency interference, realistic errors in beam models, calibration accuracy, realistic antenna position, and beam variations~\citep{orosz/etal:2019, aguirre/etal:2021, kim/etal:2022, kim/etal:2023}.

Currently, we map only 512 pixels on the sky plane, limited by the size of the $\mathbf{P}$ matrix and the calculation of the window function.
After understanding the window function, we do not need to calculate its full form; instead, we only need the sum of each $\mathbf{k}$ bin for normalization, which will dramatically reduce the RAM requirement of calculation.
The reduction will enable us to increase the size and resolution of the image cube, enlarging the $k_\bot$ range.
With an enlarged sky patch, we can include more data with a wider sky patch and develop the DOM power spectrum estimator for a curved sky~\citep{liu/zhang/parsons:2016}, which is the ultimate goal of the DOM-based power spectrum estimator.

\section{Conclusion} \label{sec:conclusion}
The image power spectrum measures $k_\parallel$ directly through sky pixels, mitigating spatial power mixing into the frequency axis.
This paper presents an FFT-based image power spectrum based on the direct optimal mapping~\citepalias{xu/etal:2022} as a first step to utilize the mapping results.

We use noiseless simulation data from \citet{aguirre/etal:2021} to explore the image power spectrum.
After obtaining the image cube with direct optimal mapping, the image cube is tapered along the three axes before performing a 3D FFT to calculate the power spectrum.
We calculate the normalization factors for the power spectrum and the power spectrum window functions.
The window functions show $<10^{-11}$ contributions from distant $k_\parallel$ power, separating the foreground from EoR signals.
The 2D and 1D power spectra further demonstrate the separation between the EoR and foreground signals.

We plan to measure the image power spectrum using HERA data~\citep{deboer/etal:2017, hera/etal:2021, hera/etal:2022, hera/etal:2024}.
The result will provide a measurement that complements the delay spectrum, with different foreground features in the 2D power spectrum.

\section*{Acknowledgements}

This analysis utilized custom-built, publicly-accessible software by the HERA Collaboration (\url{https://github.com/Hera-Team}) in addition to software built by both HERA members and collaborators (\url{https://github.com/RadioAstronomySoftwareGroup}), especially \texttt{pyuvdata} \citep{Hazelton2017}. 
This analysis also relied on number of public, open-source software packages, including \texttt{numpy} \citep{2020NumPy-Array}, \texttt{scipy} \citep{scipy2020}, \texttt{matplotlib} \citep{Hunter:2007}, and \texttt{astropy} \citep{astropy:2018}.

This material is based upon work supported by the National Science Foundation under grants \#1636646 and \#1836019 and institutional support from the HERA collaboration partners. 
This research is funded in part by the Gordon and Betty Moore Foundation through Grant GBMF5215 to the Massachusetts Institute of Technology.
HERA is hosted by the South African Radio Astronomy Observatory, which is a facility of the National Research Foundation, an agency of the Department of Science and Innovation. 
The authors wish to thank the anonymous referee for their insightful feedback.

We acknowledge the use of the Ilifu cloud computing facility (\url{www.ilifu.ac.za}) and the support from the Inter-University Institute for Data Intensive Astronomy (IDIA; \url{https://www.idia.ac.za}).
J.S.~Dillon gratefully acknowledges the support of the NSF AAPF award \#1701536.
NK acknowledges support from NASA through the NASA Hubble Fellowship grant \#HST-HF2-51533.001-A awarded by the Space Telescope Science Institute, which is operated by the Association of Universities for Research in Astronomy, Incorporated, under NASA contract NAS5-26555.
R. Byrne is supported by National Science Foundation Award No. 2303952.
P.~Kittiwisit acknowledges support from the South African Radio Astronomy Observatory (SARAO; \url{www.sarao.ac.za}) and the National Research Foundation (Grant No.\ 84156).
This result is part of a project that has received funding from the European Research Council (ERC) under the European Union's Horizon 2020 research and innovation programme (Grant agreement No.\ 948764; P.~Bull and M.J.~Wilensky). P.~Bull acknowledges support from STFC Grant ST/T000341/1.
Parts of this research were supported by the Australian Research Council Centre of Excellence for All Sky Astrophysics in 3 Dimensions (ASTRO 3D), through project number CE170100013.
G.~Bernardi acknowledges funding from the INAF PRIN-SKA 2017 project 1.05.01.88.04 (FORECaST), support from the Ministero degli Affari Esteri della Cooperazione Internazionale -- Direzione Generale per la Promozione del Sistema Paese Progetto di Grande Rilevanza ZA18GR02 and the National Research Foundation of South Africa (Grant Number 113121) as part of the ISARP RADIOSKY2020 Joint Research Scheme, from the Royal Society and the Newton Fund under grant NA150184 and from the National Research Foundation of South Africa (grant No.\ 103424).
E.~de Lera Acedo acknowledges the funding support of the UKRI Science and Technology Facilities Council SKA grant.
A.~Liu acknowledges support from the New Frontiers in Research Fund Exploration grant program, the Canadian Institute for Advanced Research (CIFAR) Azrieli Global Scholars program, a Natural Sciences and Engineering Research Council of Canada (NSERC) Discovery Grant and a Discovery Launch Supplement, the Sloan Research Fellowship, and the William Dawson Scholarship at McGill.

\bibliography{main}{}
\bibliographystyle{aasjournal}

\end{CJK*}
\end{document}